\documentclass[]{spie}  %>>> use for US letter paper
\usepackage{amsmath,amsfonts,amssymb}
\usepackage{graphicx}
\usepackage[colorlinks=true, allcolors=blue]{hyperref}
\usepackage{wrapfig}

\def\Vec#1{\mbox{\boldmath $#1$}}
\title{Statistical modeling of pneumothorax deformation by mapping CT and cone-beam CT images}
\author[a]{Megumi Nakao}
\author[a]{Hinako Maekawa}
\author[b]{Katsutaka Mineura}
\author[c]{Toyofumi F. Chen-Yoshikawa}
\author[b]{Hiroshi Date}
\author[a]{Tetsuya Matsuda}
\affil[a]{Graduate School of Informatics, Kyoto University, Kyoto, Japan}
\affil[b]{Department of Thoracic Surgery, Kyoto University Hospital, Kyoto, Japan}
\affil[b]{Department of Thoracic Surgery, Nagoya University Hospital, Nagoya, Japan}
\authorinfo{Megumi Nakao: E-mail: megumi@i.kyoto-u.ac.jp}
\begin{document} 
\maketitle
\begin{abstract}
In this study, we introduce statistical modeling methods for pneumothorax deformation using paired cone-beam computed tomography (CT) images. We designed a deformable mesh registration framework for shape changes involving non-linear deformation and rotation of the lungs. The registered meshes with local correspondences are available for both surgical guidance in thoracoscopic surgery and building statistical deformation models with inter-patient variations. In addition, a kernel-based deformation learning framework is proposed to reconstruct intraoperative deflated states of the lung from the preoperative CT models. This paper reports the findings of pneumothorax deformation and evaluation results of the kernel-based deformation framework.
\end{abstract}
\section{Introduction}
Lungs are very soft organs, and their deformation can induce considerable volume change during surgery. The position of a nodule changes because of the pneumothorax state, which makes optimization of resection procedures difficult\cite{Tokuno20}. If the pneumothorax state could be accurately estimated, precise nodule resection and preservation of pulmonary function could be facilitated by the strict management of resection margins. In the field of image-based lung modeling using computed tomography (CT) images, respiratory motion has been the main focus of investigation\cite{Ehrhardt11, Ruhaak17, Wilm16}. However, there have been few studies on the modeling of the pneumothorax deformation that occurs between the preoperative and intraoperative lung states. The mechanism is complex and not mathematically understood, except through image-based registration or modeling studies of animal lungs \cite{Nakao19, Nakao20}. As a clinical study, lung deformation generated from the posture differences between preoperative and intraoperative states was analyzed using cone-beam CT (CBCT) images \cite{Pablo18}. More recently, the deflated lung surfaces of patients with pneumothorax were evaluated using hyperelastic finite-element models \cite{Jeanne20}. In our previous study, we acquired a pair of CBCT images in the inflated and deflated states during thoracoscopic surgery and obtained registration results of the pneumothorax-associated deformation\cite{Maekawa20}.

In this study, we introduce statistical modeling methods for pneumothorax deformation using paired CBCT images. We designed a deformable mesh registration (DMR) framework for shape changes involving non-linear deformation and rotation. The registered meshes with local correspondences are directly available for surgical guidance in thoracoscopic surgery as well as for building statistical deformation models with inter-patient variations. In addition, a kernel-based deformation learning framework is proposed to reconstruct intraoperative deflated states of the lung from the preoperative CT models. This paper reports the findings of pneumothorax deformation and evaluation results of the kernel-based deformation framework.

\section{Methods}
\subsection{Paired CBCT images}
Intraoperative CBCT images were acquired from twelve lung cancer patients at Kyoto University Hospital. All participants provided informed consent before their enrollment and data collection. The analyses were approved by the institutional review board of Kyoto University Hospital. Two CBCT image sets of the inflated and deflated states were measured in the same lateral position by controlling the bronchial pressure during thoracoscopic surgery. The images were measured with two surgical clips placed on the lung surface near the tumor as physical landmarks. Each CBCT volume consists of $512 \times 512$ pixel image slices (voxel resolution: $0.49$ mm $\times 0.49$ mm $\times 0.49-1.0$ mm). 

\begin{figure}[h]
  \centering
  \includegraphics[width=105mm]{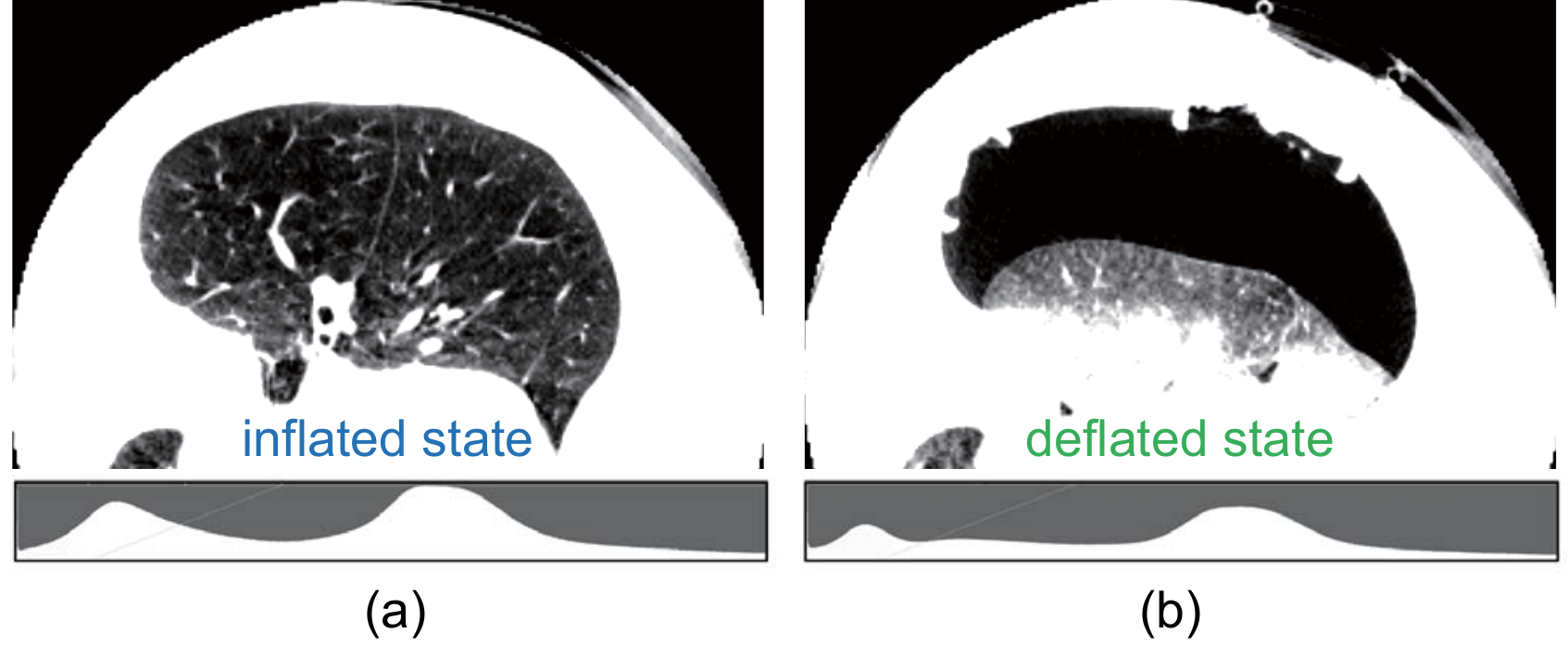}
  \caption{Paired CBCT images and histogram of CT values. (a) Inflated and (b) deflated states.
}
  \label{fig:1}
\end{figure}

Figure \ref{fig:1} shows the CBCT slices after registering the two volumes using the spine as a fixed reference. A large space caused by air flowing into the thoracic cavity in the deflated state is confirmed. Interestingly, the volume of the lung in the field of view becomes less than a third of the inflated volume, and the posture greatly influences the deformation. The CT values change because of differences in the air content of the lung, and the parenchyma region becomes brighter in the deflated state. These image features are consistent with the features in CT images measured from beagle dogs \cite{Nakao19} in our past study. Because conventional image-based registration methods did not work correctly given the characteristics of these images (i.e., CT value shifts and topological changes due to the thoracic cavity space), we applied deformable mesh registration (DMR) to partial lung meshes created from paired CBCT images.

\subsection{DMR for pneumothorax deformation}
Three-dimensional labeling of lungs, bronchi, and tumors was performed from CBCT images using Synapse VINCENT (Fujifilm Corporation, Japan), and tetrahedral meshes of lungs with 500 vertices were generated. The bronchial structures were extracted from the inflated lungs because of the low image contrast in the images of the deflated states. This means that only the lung's surface and the placed surgical clips were available for registering the inflated/deflated lungs. The number of vertices was determined to balance the need for accuracy against the calculation time. 

DMR was performed from the source mesh (inflated state) to the target shape (deflated state). To achieve both globally stable and locally strict DMR, our framework addresses the trade-off problem between feature-preserving shape matching and spatially smooth deformation\cite{Nakao21}. Because registration errors tend to increase in cases of large deformation with rotation, we use the position of the two surgical clips as additional constraints. The objective function is described as follows.
\begin{align}
    E(\Vec{u}) &= E_{shape} + E_{clip} \\ \nonumber
    &= d(S^D, \phi(S^I)) + \omega \| \Vec{p}^D - \Vec{p}^I \|_2,  
\end{align}
where $S^I$ is the source mesh (inflated state), $S_D$ is the target mesh (deflated state), and $d$ is the mean distance, which is mean value of the nearest bidirectional point-to-surface distance of the two meshes. Moreover, $\phi(X)$ is a continuous and differentiable transformation that maps $X$ to the deformed mesh, $w$ is a weight, and $\Vec{p}^D$ and $\Vec{p}^I$ are the positions of the two surgical clips placed on the lung surface in the inflated and deflated states, respectively. The registered (deformed) mesh is obtained by minimizing the cost function in Eq. (1). The first term evaluates the difference between the deformed source and the target mesh. It also enforces the preservation of the discrete Laplacian, a shape descriptor that approximates the mean curvature normal of the mesh \cite{Nakao20}.
\begin{figure}[t]
  \centering
  \includegraphics[width=135mm]{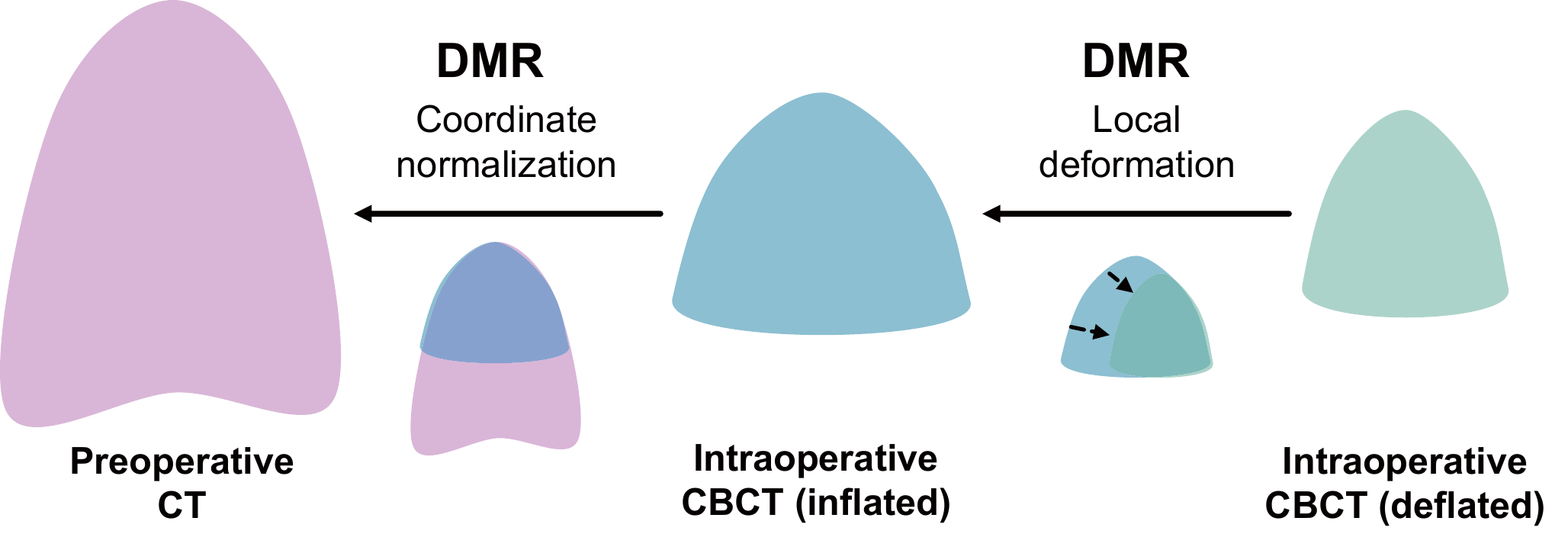}
  \caption{Statistical modeling framework for pneumothorax deformation, showing inter- and intra-patient mapping between preoperative CT and paired CBCT images.}
  \label{fig:2}
\end{figure}
In addition to intra-patient registration of the CBCT images, our framework is designed to build a statistical model for pneumothorax deformation, making inter-patient deformation analysis possible. However, the entire lung cannot be measured because of the limited field of view in CBCT imaging. To address this issue, we introduce the concept of statistical modeling into our approach by synthesizing partial CBCT models of patients (see Figure \ref{fig:2}). In this framework, the local deformation or the spatial displacement map is captured from the paired CBCT models. The registered CBCT models are mapped onto a unique coordinate system using the preoperative CT image. This concept makes it possible to evaluate local shapes and deformations in the global coordinate while normalizing the coordinates. In addition, a statistical model that represents the inter-patient variation of both lung geometry and pneumothorax-associated deformation can be built by registering the preoperative CT models between patients. 

\subsection{Localized kernel for deformation reconstruction}
\begin{wrapfigure}[10]{r}[5mm]{85mm}
  \centering
  \includegraphics[width=80mm]{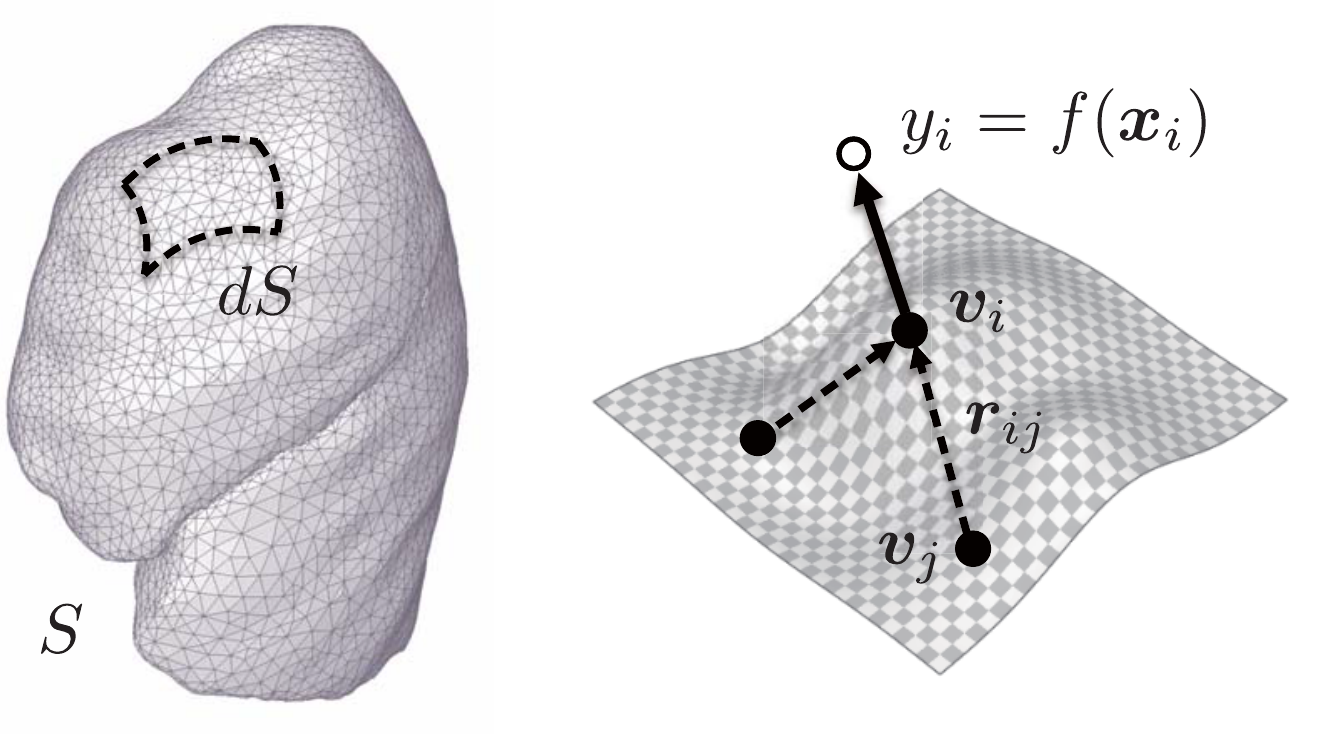}
  \caption{Per-region-based deformation learning model using kernel functions.}
  \label{fig:3}
\end{wrapfigure}

In this study, we aim to reconstruct pneumothorax deformation and to learn the mapping from the registered lung meshes. For this purpose, we introduce the per-region-based deformation learning model using kernel functions \cite{Nakao21}. Per-patient-based learning is a straightforward approach in which the mesh model $S$ obtained from one patient’s data is used as one set of training data. This approach is based on the idea that the displacement in the local regions of organs is similar to that of the corresponding regions of the training dataset with similar shape features. However, in the proposed per-region-based localized deformation learning, a small region $dS$ of mesh model $S$ is used as one set of training data. This approach is also based on the physical characteristics of deformation in that two local regions that are close to each other have a similar displacement because the target surface is generally assumed to be a smooth curved manifold (see Figure \ref{fig:3}).

In the per-region-based deformation learning model, we hypothesize that local displacement $\Vec{y}_i$ can be calculated from feature vectors $\Vec{x}_i$, which are defined by the relative position $\Vec{r}_{ij}$ of the sampled vertex, using Eq. (2). 
\begin{align}
\Vec{y}_{i} = \sum_{j=1}^{N} \Vec{\alpha}_j k(\Vec{x}_{i},\Vec{x}_j), \hspace{.3cm} \Vec{x}_{j}\in \it{X}, \hspace{.3cm}\Vec{\alpha}_{j}\in\mathbb{R}^{N},
\label{eq:kernel_regression}
\end{align}
where $k :\it{X} \times \it{X}\rightarrow\mathbb{R}$ is the kernel function defined for scalar component of the three-dimensional displacement vector, $N$ is the number of training datasets, and $\Vec{\alpha}_{j}$ is the weight vector. A Gaussian function is used for kernel function $k$, which is $k(\Vec{x}_i,\Vec{x}_j) = exp(- \beta ||\Vec{x}_i - \Vec{x}_j||^2 / N)$. For a given $\Vec y = [\Vec{y}_1, \dots, \Vec{y}_N]^T$, $\Vec \alpha = [\Vec{\alpha}_1, \dots,\Vec{\alpha}_N]^T$ are calculated by minimizing the cost function E(\Vec{\alpha}), which is expressed as
\begin{align}
E(\Vec{\alpha})	= \| \Vec{y} - K \Vec{\alpha} \|^{2} + \lambda \Vec{\alpha}^{T} K \Vec{\alpha},
\label{eq:kernel_cost_function}
\end{align}
where $K \in \mathbb{R}^{N \times N} $ is the kernel matrix, whose elements are defined by $K_{ij} = k(\Vec{x}_{i},\Vec{x}_j)$, and $\lambda$ is the regularization parameter, which penalizes deviations of $\Vec{\alpha}$. The optimized weights are given by $\Vec \alpha = (K + \lambda I)^{-1} \Vec y$ ($I$: identity matrix). The feature vector $\Vec{x}_{i}$ of the local region for per-region-based learning is constructed using the relative position $\Vec{r}_{ij}$ of $n$ sampled vertices as follows.
\begin{align}
\Vec{x}_{i} = \{\Vec{r}_{i1}, \Vec{r}_{i2}, \dots, \Vec{r}_{in}\}, \hspace{.3cm} \Vec{r}_{ij} = \Vec{v}_i - \Vec{v}_j.
\label{eq:feature_vector}
\end{align}
In this scheme, the displacement is locally learned per vertex; in other words, it can be reconstructed from the deformation of different regions. Therefore, using a per-region-based kernel formulation improves the estimation performance and more stable results can be expected even for a small number of patient datasets\cite{Nakao21}. 

\section{Results}
In the experiments, the registration performance of twelve CBCT datasets was confirmed by comparing the two methods with and without surgical clips for the constraints. The pneumothorax deformation was analyzed using the registered results, and preliminary results on the prediction performance of the kernel-based deformation framework were obtained. We used 0.1 for the hyperparameter $\lambda$ in Eq. (3) after an examination of several parameters sets. 

\subsection{Registration accuracy}
The mean distance (MD), Hausdorff distance (HD)\cite{Nakao19}, and target registration error (TRE) of the two surgical clips were used as the evaluation criteria. The HD measures the longest distance between two surfaces, whereas the MD is the mean value of the nearest bidirectional point-to-surface distance\cite{Nakao19}. Table \ref{table:1} presents the registration performance. There are no significant differences in either distance metric, MD or HD (one-way analysis of variance, ANOVA; $p < 0.05$ significance level). In contrast, both TRE1 and TRE2 have significantly smaller errors of less than 5 mm when surgical clips are used.

\begin{table}[b]
    \centering
    \caption{Quantitative comparison of mean (minimum--maximum) registration errors. Mean distance (MD), Hausdorff distance (HD) and target registration errors (TRE) are listed.}
    \begin{tabular}{c c c}
    	\hline
    	&no clips&with clips\\\hline
    	MD [mm]&0.1 (0.1 - 0.2)&0.1 (0.1 - 0.2)\\
    	HD [mm]&1.8 (0.8 - 4.6)&1.5 (0.7 - 3.2)\\
    	TRE1 [mm]&20.4 (7.5 - 39.3)&3.9 (1.8 - 6.8)\\
    	TRE2 [mm]&21.4 (7.9 - 35.8)&4.0 (2.2 - 5.8)\\
		\hline
	\end{tabular}    
    \label{table:1}
\end{table}

\begin{figure}[t]
  \centering
  \includegraphics[width=150mm]{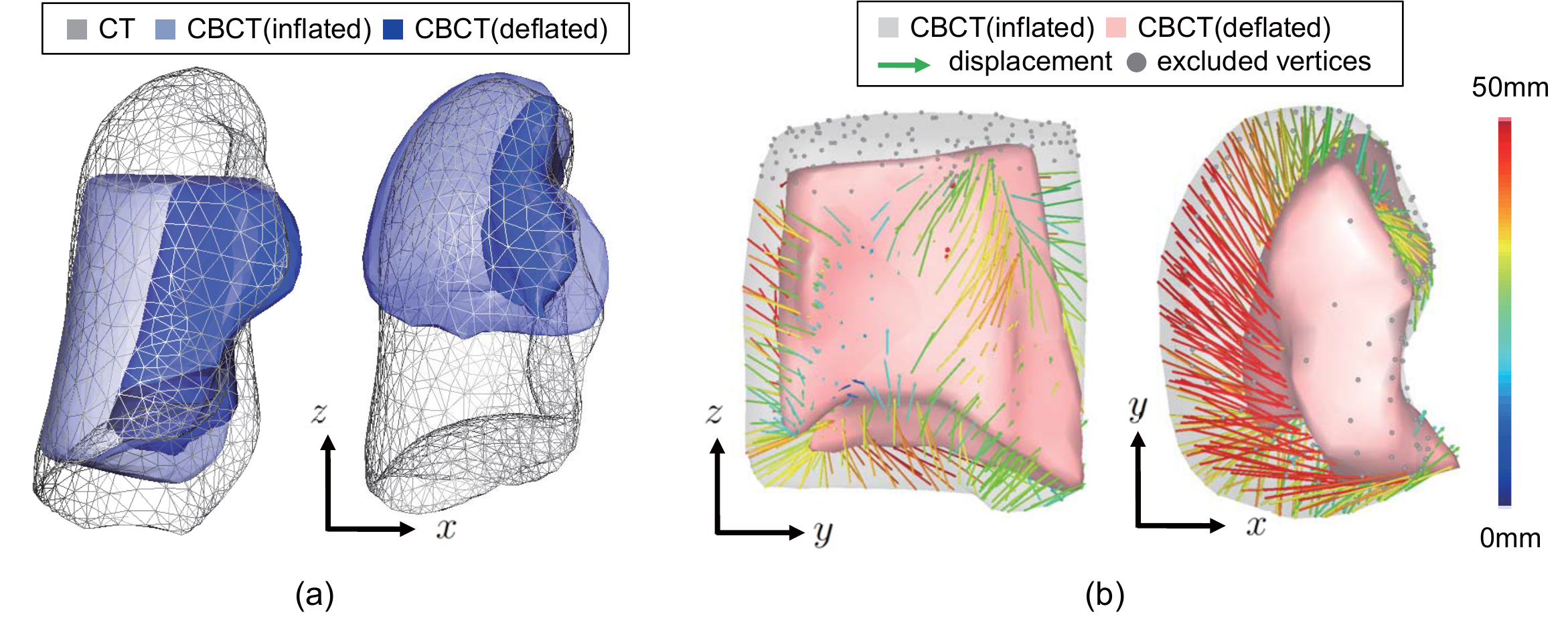}
  \caption{Deformable mesh registration results. (a) Inflated/deflated CBCT meshes registered in CT coordinates and (b) spatial displacement vectors obtained from the corresponding vertices.}
  \label{fig:4}
\end{figure}

Figure \ref{fig:4} (a) shows the registered results of the preoperative CT and intraoperative CBCT mesh models acquired from inflated and deflated states of the lung, and Figure \ref{fig:4} (b) illustrates the spatial displacement vectors obtained from corresponding vertices of the registered mesh models. Interestingly, it was found that in addition to contraction, the lung surface deformation includes rotation. We believe that this was due to gravitational force, and the TREs of the conventional DMR algorithm without landmarks significantly increased because the algorithm could not cope with a non-linear deformation that includes rotation. 

\subsection{Deformation reconstruction performance}
\begin{figure}[t]
  \centering
  \includegraphics[width=150mm]{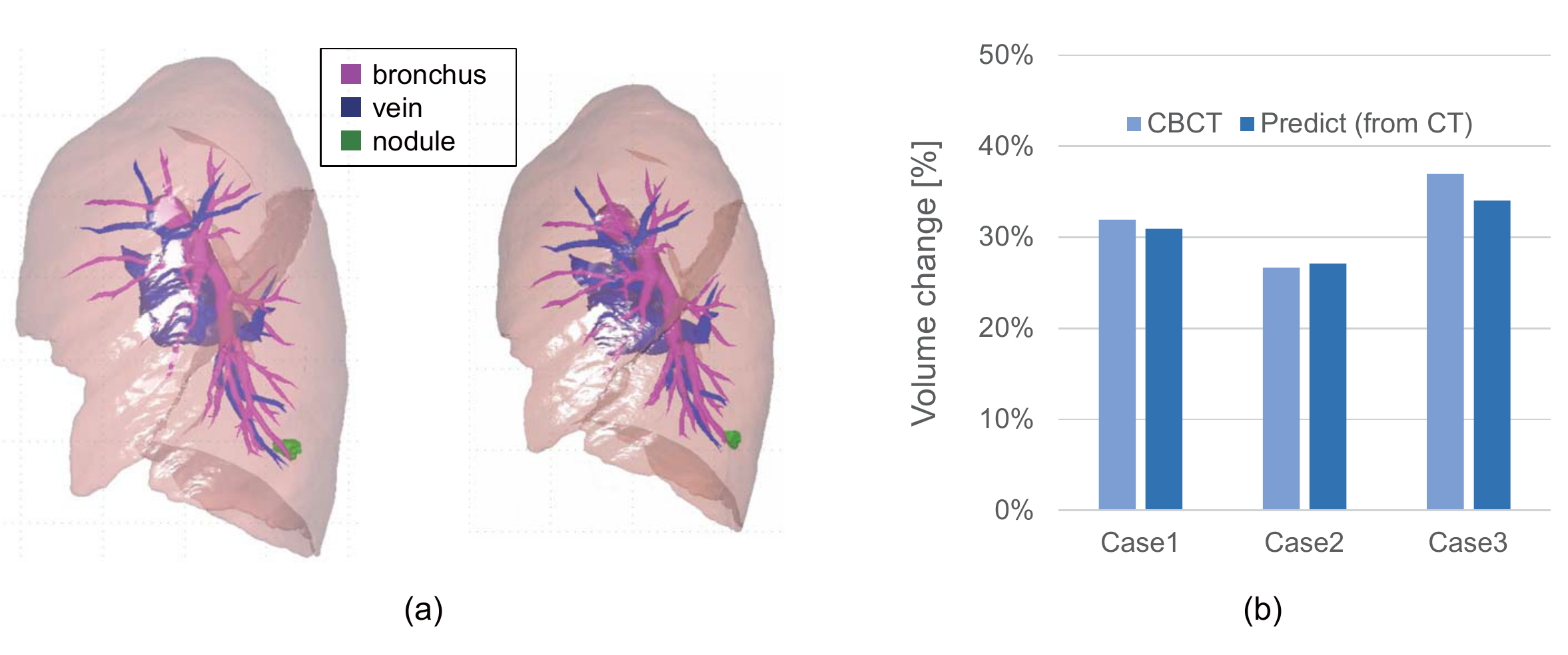}
  \caption{Kernel-based deformation reconstruction from preoperative CT model. (a) Preoperative CT model (left) and estimation result of the intraoperative pneumothorax state (right). (b) Comparison of the volume change ratio between the paired CBCT model and predicted model.}
  \label{fig:5}
\end{figure}

The second experiment was conducted to investigate the prediction performance of the per-region-based learning model. The left image in Figure \ref{fig:5} (a) shows the initial shape with the tumor and vascular structures generated from the CT images, and the right image is the estimated result of the intraoperative deflated state. The proposed kernel model could represent the non-linear pneumothorax-associated deformation with rotation. Subsurface structures were visualized with a linear mapping, that is, the displacements for the lung nodule, bronchi, and vascular structures were linearly interpolated from the deformed surfaces.
To confirm the prediction performance, we conducted a leave-one-out cross-validation, where the deflated state was reconstructed from the inflated state of the preoperative CT models for the left lung. Figure \ref{fig:5} (b) shows the volume changes in the three test cases along with the ones in the CBCT images. The preliminary results showed that the average estimation error was 4.0\% in volume change and 5.1 mm in the TRE of the surgical clips. A limitation of the current experiments is that only 12 subjects provided the image data. However, despite the amount of data, these results show that the proposed framework has potential for estimating pneumothorax deformation from preoperative CT images, which will contribute to nodule localization during thoracoscopic surgery.

\section{Conclusion}
In this paper, we introduced data-driven modeling methods for pneumothorax deformation using paired CBCT images. Inter- and intra-patient DMR of the CT and CBCT models enabled the statistical representation of intraoperative pneumothorax states. The kernel-based learning framework was able to reconstruct intraoperative deflated states with a clinically acceptable estimation error.

\acknowledgments
This research was funded by the Medical Arts Program from the Japan Agency for Medical Research (AMED). A part of this study was also supported by a JSPS Grant-in-Aid for challenging Exploratory Research (grant number 18K19918). We thank Kimberly Moravec, PhD, from Edanz Group (https://en-author-services.edanz.com/ac) for editing a draft of this manuscript.

\bibliographystyle{IEEEtran}

%\bibliography{report} % bibliography data in report.bib
%\bibliographystyle{spiebib} % makes bibtex use spiebib.bst
\end{document}